\documentclass[12pt,preprint]{aastex}
\usepackage{natbib}

\slugcomment{NAOJ-Th-Ap 2002,No.17, ApJ in press}

\shorttitle{Subhalos and Cold Fronts in Clusters}
\shortauthors{Fujita, et al.}

\begin{document}

\title{Analytical Approach to the Mass Distribution Function of Subhalos
and Cold Fronts in Galaxy Clusters}

\author{Yutaka Fujita\altaffilmark{1,2}, Craig
L. Sarazin\altaffilmark{2}, 
Masahiro Nagashima\altaffilmark{1}, and Taihei Yano\altaffilmark{1}}

\email{yfujita@th.nao.ac.jp}

\altaffiltext{1}{National Astronomical Observatory, Osawa 2-21-1,
Mitaka, Tokyo 181-8588, Japan} 
\altaffiltext{2}{Department of Astronomy,
University of Virginia, P. O. Box 3818, Charlottesville, VA 22903-0818}

\begin{abstract}
 We construct an analytical model of the mass distribution function of
 subhalos in a galaxy cluster in the context of the cold dark matter
 (CDM) theory. Our model takes account of two important effects; the
 high density of the precluster region and the spatial correlation of
 initial density fluctuations. For subhalos with small masses, the mass
 distribution function that our model predicts is intermediate between
 the Press-Schechter (PS) mass distribution function and a conditional
 mass distribution function as an extension of the PS formalism. We
 compare the results of our model with those of numerical
 simulations. We find that our model predictions are consistent with the
 mass and velocity distribution functions of subhalos in a cluster
 obtained by numerical simulations. We estimate the probability of
 finding large X-ray subhalos that often have ``cold fronts''.  The
 observed large X-ray subhalos and cold fronts may not always be the
 result of cluster mergers, but instead may be internal structures in
 clusters.
\end{abstract}


\section{Introduction}

It is generally believed that dark matter constitutes a large fraction
of the mass in the universe. Among various theories of dark matter, cold
dark matter (CDM) theory provides a remarkably successful description of
large-scale structure formation, and it is in good agreement with a
large variety of observational data. This model predicts that small
objects are the first to form and these then amalgamate into
progressively larger system. This means that galaxy clusters have
formed via merging and accretion of galaxies or galaxy groups.

If the dark matter is actually the main component of galaxies and
clusters, $N$-body simulations can be used to predict their bulk
properties, such as position, mass, and size.  However, it has proven
difficult to find surviving substructure (subhalos) in virialized
objects in $N$-body simulations.  This apparent absence of substructure,
known as ``the overmerging problem,'' reflects the fact that simulated
subhalos are disrupted much more efficiently than real subhalos
\citep*{moo96}.

Analytical studies showed that the overmerging problem was caused by
poor force and mass resolutions \citep{moo96}. \citet{kly99} estimated
that the force and mass resolutions required for a simulated halo to
survive in galaxy groups and clusters are extremely high: $\sim 1-3$~kpc
and $10^8-10^9\; M_{\sun}$. Recently, simulations with roughly the
required resolution have been done
\citep*{ghi98,tor98,kly99,oka99,col99,moo99,jin00,ghi00,got01,whi01,
fuk01,bul01}. These simulations certainly showed that subhalos survive
in clusters. \citet{oka99} compared the mass distribution function (MDF)
of subhalos with the Press-Schechter (PS) MDF \citep{pre74} and a
conditional MDF derived through an extension of the PS formalism
\citep[EPS;][see \S\ref{sec:MDF} for detail]{bow91,bon91,lac93}. They
found that the MDF of subhalos in their simulated cluster at $z=0$ is
similar to the PS MDF at $z=2$. It is also similar to the EPS MDF but
the fit is a little worse. They suggested that the MDF of subhalos
freezes out after the cluster had grown sufficiently, which in their
simulation occurred at $z\sim 2$.  Thus, the MDF of subhalos preserves
the information about the growth of density fluctuations at that time.
In other words, we can use observations of the MDF of subhalos at
present to determine the hierarchical clustering which occurred in the
past.  However, the good match between the simulations and the PS MDF
found by \citet{oka99} is somewhat surprising as the PS MDF does not
take into account the fact that the cluster subhalos are in a high
density region.  The EPS MDF does include the density effect, yet
provides a slightly worse fit to the simulations than the PS MDF.

Another important factor that should be taken into account when the MDF
of subhalos in a cluster is considered is the spatial correlation of
initial density fluctuations. It has been shown that this spatial
correlation does affect the MDF of dark halos in the whole universe
\citep*{yan96, nag01}. This means that one cannot ignore the effect of
spatial correlation on the MDF of subhalos because these subhalos
originated from density fluctuations that were close to one another. In
this paper, we construct a model of the subhalo MDF in a cluster
considering the effect of spatial correlations as well as the fact that
cluster subhalos are located in a high density region. Of course, there
are other mechanisms that could affect the MDF of subhalos especially
after the cluster formation (e.g. tidal stripping).  Although they
should be taken into account in order to derive the MDF exactly, we do
not consider them in our model for the sake of simplicity. Even so, our
model should be useful to predict the MDF in the case where the
mechanisms that occur after cluster formation are not very effective,
and gives an upper limit to substructure if disruptive processes are
important for a wide range of masses.  Our analytic model is not
affected by spatial and mass resolutions, which affect numerical
simulations.  It will be useful to compare with the results of numerical
simulations to determine the effects of finite resolution.  Even though
some of the recent high-resolution simulations meet fully the conditions
required to study subhalos \citep{kly99}, comparing them with our model
is still important.  Such comparisons can serve as a check for the
subhalo identification algorithm applied to the simulation results.

Both the MDFs predicted by our model and the results from numerical
simulations should be compared with the observations.  In the past,
galaxies were used to derive the observed MDF in a cluster.  In fact,
\citet{moo99} compared their simulation results with the observational
data from the Virgo cluster, although they used the circular velocity
distribution function (VDF) of galaxies instead of the MDF.  They showed
that the simulation and observed VDFs are roughly consistent with one
another.  However, individual galaxies could not be used to detect
massive subhalos, because the observed circular velocity (or velocity
dispersion) of a galaxy is at most $v_c \sim 300\;\rm km\; s^{-1}$. On
the other hand, numerical simulations suggest that clusters have
subhalos with $v_c \gtrsim 300\;\rm km\; s^{-1}$ \citep{moo99}. Until
recently, it was not understood whether the lack of galaxies with $v_c
\gtrsim 300\;\rm km\; s^{-1}$ requires a lack of massive subhalos in a
cluster, or whether this implies that galaxies have lower values of
$v_c$ than would characterize the massive dark matter halos surrounding
the galaxies.  Recently, X-ray observations with the {\it Chandra}
Observatory have found that some clusters have massive subhalos moving
within the clusters which have not been disrupted
\citep*{mar00,vik01,maz01,maz02,sun02}. The contact surface between the
cooler subhalo gas and the surrounding hot intracluster medium is called
a ``cold front''. The X-ray temperatures of these subhalos are $\sim
4-7$~keV, which is substantially higher that the kinetic temperatures
one would derive from the motions of stars in the optical galaxies in
the subhalos.  The observations of these massive subhalos gives us
important information about the massive end of the MDF.

The plan of our paper is as follows.  In \S\ref{sec:model}, we derive
the MDF in clusters analytically.  We discuss the resulting MDFs and
compare them with those from numerical simulations in \S\ref{sec:MDF},
We discuss the effect of tidal stripping and dynamical friction on the
MDFs in \S\ref{sec:tidal}. In \S\ref{sec:VDF}, we derive the VDF of
subhalos and compare these results with those of numerical simulations
and observations of galaxies.  We estimate the probability of finding
massive X-ray subhalos and associated cold fronts in clusters in
\S\ref{sec:coldf}.  Finally, \S\ref{sec:conc} summarizes our
conclusions.

\section{Models}
\label{sec:model}

In this section, we derive the MDF of subhalos in a cluster including
the effects of spatial correlations using the model of \citet{yan96}. We
use the so-called sharp $k$-space filter for density fluctuations
following \citet{yan96}. We assume that the primordial density field is 
a Gaussian random field.

First, we estimate the conditional probability, $P(r,M_1,M_2)$, of
finding a region of mass $M_1$ with $\delta_1\geq\delta_{c1}$ at a
distance $r$ from the center of an isolated, finite-sized object of mass
$M_2$, provided that the object of mass $M_1$ is included in the object
of mass $M_2$ $(>M_1)$ with $\delta_2=\delta_{c2}$ at $r=0$.  Here, we
define $\delta_M$ as the smoothed linear density fluctuation of mass
scale $M$, and $\delta_1=\delta_{M_1}$ and $\delta_2=\delta_{M_2}$.
Moreover, we define $\delta_c(z)$ as the critical density threshold for
a spherical perturbation to collapse by the redshift $z$, and
$\delta_{c1}=\delta_{c}(z_1)$ and $\delta_{c2}=\delta_{c}(z_2)$.  In the
Einstein-de Sitter universe, $\delta_c(z)=1.69 (1+z)$. While
\citet{yan96} considered only the case of $\delta_{c1}=\delta_{c2}$, we
consider the case of $\delta_{c1}\geq \delta_{c2}$ ($z_1>z_2$) in
general. Because the covariance matrix for the Gaussian distribution for
two variables $[\delta_1(r), \delta_2]$ is given by
\begin{equation}
\mbox{\boldmath $M$}= \left[
\begin{array}{cc}
\sigma_1^2 & \sigma_c^2(r) \\
\sigma_c^2(r) & \sigma_2^2 \\
\end{array}
\right]\;,
\end{equation}
the probability can be written as 
\begin{eqnarray}
 P(r,M_1,M_2) 
  &=& p[\delta_1\geq\delta_{c1}| \delta_2=\delta_{c2}] \nonumber \\
  &=& p[\delta_1\geq\delta_{c1}, \delta_2=\delta_{c2}]
      /p(\delta_2=\delta_{c2}) \nonumber \\
  &=& \frac{1}{\sqrt{2\pi [1-\epsilon^2(r)]}}
      \int^{\infty}_{\nu_{1c}}
      \exp{\left\{\frac{-[\nu_1-\epsilon(r)\nu_{2c}]^2}
      {2[1-\epsilon^2(r)]}\right\}}d\nu_1\;, \label{eq:probr}
\end{eqnarray}
where $\nu_1$ and $\nu_2$ are defined by
\begin{equation}
\label{eq:nu}
 \nu_1 \equiv \frac{\delta_1}{\sigma_1} \, , \,
 \nu_2 \equiv \frac{\delta_2}{\sigma_2} \, , \,
 \sigma_1 \equiv \sigma(M_1) \, , \,
 \sigma_2 \equiv \sigma(M_2) \, , \,
 \nu_{1c} \equiv \frac{\delta_{c1}}{\sigma_1} \, , \,
 \nu_{2c} \equiv \frac{\delta_{c2}}{\sigma_2} \, , \,
\end{equation}
respectively \citep{yan96}. In equations (\ref{eq:nu}), $\sigma(M)$ is
the rms density fluctuation smoothed over a region of mass $M$,
and is given by
\begin{equation}
 \sigma^2(M)=\frac{V}{(2\pi)^3}\int_0^{k_c(M)}|\delta_k|^2
 4\pi k^2 dk\;.
\end{equation}
In this equation, $V$ is the volume, $k$ is the wave number,
and $\delta_k$ is the Fourier components of density fluctuations.  The
critical wave number, $k_c(M)$ is the wave number corresponding to mass
scale $M$ and is given by $k_c(M)=\pi/(3M/4\pi\bar{\rho})$, where
$\bar{\rho}$ is the current mean mass density of the universe.
Moreover, $\epsilon(r)$ is defined by
\begin{equation}
 \epsilon(r) \equiv \frac{\sigma_c^2(r)}{\sigma_1 \sigma_2}\;,
\end{equation}
\begin{equation}
\label{eq:two}
 \sigma_c^2(r)=\langle\delta_{2}(\mbox{\boldmath $r$}_0)
\delta_{1}(\mbox{\boldmath $r$}_0+\mbox{\boldmath $r$})\rangle
=\frac{V}{(2\pi)^3}\int_0^{k_c(M_2)}|\delta_k|^2\frac{\sin(kr)}{kr}
 4\pi k^2 dk\;,
\end{equation}
which corresponds to the two-point correlation function.

We can rewrite equation (\ref{eq:probr}) as
\begin{equation}
\label{eq:probr1}
 P(r,M_1,M_2) =\frac{1}{\sqrt{2\pi}}\int_{\beta}^{\infty} 
             e^{-y^2/2} dy \;,
\end{equation}
where
\begin{equation}
\label{eq:beta}
 \beta(r)=\frac{\nu_{1c}-\epsilon(r)\nu_{2c}}{\sqrt{1-\epsilon^2(r)}}
      = \frac{1}{\sqrt{1-\epsilon^2(0)\alpha^2(r)}}
         \frac{\delta_{c1}}{\sigma_1}
\left[1-\frac{\delta_{c2}}{\delta_{c1}}\alpha(r)\right]\:,
\end{equation}
and $\alpha(r)=\epsilon(r)/\epsilon(0)$.  The spatially averaged
conditional probability for $r<R$ is defined as
\begin{equation}
\label{eq:prob}
 P(M_1, M_2)=\int_0^R P(r, M_1, M_2) 4\pi r^2 dr/\int_0^R 4\pi r^2 dr\:.
\end{equation}

The definition of $P(M_1, M_2)$ (equations [\ref{eq:probr1}] and
[\ref{eq:prob}]) is similar to the probability that the
density contrast on a scale of $M$ exceeds $\delta_c$ in field:
\begin{equation}
\label{eq:fPS}
 f(\geq\delta_c, M)=\frac{1}{\sqrt{2\pi}}
\int_{\delta_c/\sigma(M)}^{\infty} 
             e^{-y^2/2} dy \;.
\end{equation}
 From this, the PS MDF can be obtained as
\begin{equation}
\label{eq:PS}
 \tilde{n}_{\rm PS}dM =2\frac{\bar{\rho}}{M}
\left|\frac{df}{dM}\right|dM \:,
\end{equation}
\citep{pre74}.

Using the similarity between $f(\geq\delta_c, M)$ and $P(r,M_1,M_2)$, we
define the MDF of the collapsed objects (subhalos) in the region of a
mass scale of $M_0$ by differentiating $P(M, M_0)$ and multiplying it by
2 (PS approximation):
\begin{equation}
\label{eq:yng}
 n_{\rm SPS}dM =2 \frac{M_0}{M}\left|\frac{\partial P(M, M_0)}
{\partial M}\right|dM \:.
\end{equation}
This expression is based on the usual PS assumptions.  However, we
emphasize that the MDF (equation~[\ref{eq:yng}]) includes the effect of
the spatial correlation of fluctuations through the two-point
correlation function (equation~[\ref{eq:two}]). Moreover, it is the MDF
in a high density region, rather than the MDF for an average region of
the universe.  As will shown in the next section, our model is a further
extension of the EPS MDF by taking account of spatial correlations of
halos. Thus, we refer to this MDF as SPS MDF. We assume that the region
of the mass scale of $M_0$ will collapse into a cluster and the subhalos
in the region also fall into the cluster. The density fluctuations in
the cluster region should stop growing as the cluster grows, because the
kinetic energy released at the cluster collapse allows the subhalos to
move in the cluster, which prevents their further growth. We call the
redshift when this occurs the effective formation redshift of the
cluster, $z_f$. Thus, we use $\delta_{c1}=\delta_c(z_f)$ or $z_1=z_f$ in
equations (\ref{eq:probr}) and (\ref{eq:yng}).

Strictly speaking, the conditional probability $P(M_1,M_2)$ derived in
this section is not sufficient to obtain the exact MDF. This is because
an isolated object of mass scale $M_2$ must have the {\it maximum peak
density} with $\delta_2=\delta_{c2}$ at $r=0$.  Thus, we need an
additional condition. However, \citet{yan96} showed that the effect is
negligible as long as we consider a fluctuation power spectrum with an
index of $n\lesssim -2$ and the mass scale with $\sigma(M)\gtrsim
1$. Our calculations in the following sections satisfy these conditions.

\section{Mass Distribution Function}
\label{sec:MDF}

We calculate the MDF of subhalos in a cluster, $n_{\rm SPS}$, for a CDM
universe with cosmological parameters of $\Omega_0=1$, $\lambda=0$,
$\sigma_8=0.7$, $\Gamma=0.5$, and $h=0.5$, where $\sigma_8$ is the rms
mass fluctuation on $8 h^{-1}$~Mpc, and $\Gamma$ is the so-called shape
parameter of the CDM spectrum. We use a Hubble constant of $H_0=100
h\;\rm km\;s^{-1}\; Mpc^{-1}$. The cluster mass at $z_0=0$ is
$M_0=4.3\times 10^{14}\; M_{\sun}$.  These parameters are the same as
those used for an ultra-high resolution numerical simulation done by
\citet{ghi00}. We calculate the conditional probability, $P(M,M_0)$,
spatially averaged in a precluster region, $r<R=R_0$, where
$R_0=(3M_0/4\pi\bar{\rho})^{1/3}$.  From now on, we assume $R=R_0$
unless otherwise stated.

For comparison, we also calculated a conditional MDF, that is, the
number of halos with mass between $M$ and $M+dM$ at $z_f$ that are in a
halo of mass $M_0$ at $z_0$ ($M<M_0$, $z_0<z_f$). It is given by
\begin{equation}
\label{eq:cond}
 n_{\rm EPS}(M, z_f|M_0, z_0)dM=\frac{M_0}{M}f(S,\omega|S_0,\omega_0)
\left|\frac{dS}{dM}\right|dM \:,
\end{equation}
where
\begin{equation}
 f(S,\omega|S_0,\omega_0)dS = 
\frac{\omega-\omega_0}{(2\pi)^{1/2}(S-S_0)^{3/2}}
\exp\left[-\frac{(\omega-\omega_0)^2}{2(S-S_0)}\right]dS \;,
\end{equation}
$S=\sigma^2(M)$, $S_0=\sigma^2(M_0)$, $\omega=\delta_c(z_f)$, and
$\omega_0=\delta(z_0)$ \citep{bow91,bon91,lac93}. This formalism is
often called the extended Press-Schechter formalism (EPS).

We note that the difference between $n_{\rm SPS}$ and $n_{\rm EPS}$ comes
from whether the effect of the spatial correlation of fluctuations is
included or not. If we ignore the spatial correlation, $P(M_1,M_2)$
should be written as
\begin{eqnarray}
\label{eq:nocorr}
 P(M_1,M_2) 
&=& p(\delta_1\geq \delta_{c1}|\delta_2=\delta_{c2})\nonumber \\
&=& \int_{\delta_{c1}}^{\infty}
 \frac{1}{\sqrt{2\pi (\sigma_1^2-\sigma_2^2)}} 
 \exp\left[-\frac{(\delta_1-\delta_{2c})^2}
 {2(\sigma_1^2-\sigma_2^2)}\right]d\delta_1 \nonumber \\
&=& \frac{1}{\sqrt{2\pi}}\int_{\beta'}^{\infty} 
             e^{-y^2/2} dy \;, 
\; 
\end{eqnarray}
where
\begin{equation}
\label{eq:beta'}
 \beta'= \frac{1}{\sqrt{1-\epsilon^2(0)}}
         \frac{\delta_{c1}}{\sigma_1}
\left(1-\frac{\delta_{c2}}{\delta_{c1}}\right)\:,
\end{equation}
\citep[see][]{yan96}. In this case, it can easily be shown that the
right hand of equation (\ref{eq:yng}) is identical to that of equation
(\ref{eq:cond}) \citep[see][]{bow91}. We also note that at $r=0$, the
right hand of equation (\ref{eq:probr}) is identical to that of equation
(\ref{eq:nocorr}) \citep{yan96}. We refer the right hand of equation
(\ref{eq:nocorr}) as $P_{\rm EPS}(M_1,M_2)$, so that
\begin{equation}
\label{eq:c}
 n_{\rm EPS}\: dM =2 \frac{M_0}{M}\left|\frac{\partial 
P_{\rm EPS}(M, M_0)}
{\partial M}\right|dM \:.
\end{equation}
We would like to point out that the equation (\ref{eq:prob}) gives the
probability distribution of the mass of halo progenitors as $P_{\rm
EPS}(M_1, M_2)$. Thus, it could be possible to construct dark halo
merger trees including the spatial correlation of dark halos, which
would be important for semi-analytic models of galaxy formation.

We also compare $n_{\rm SPS}$ with the conventional PS MDF (equation
[\ref{eq:PS}]). Instead of $\tilde{n}_{\rm PS}$, we use the MDF,
\begin{equation}
\label{eq:PS2}
 n_{\rm PS}dM = \frac{M_0}{\bar{\rho}} \tilde{n}_{\rm PS}dM
=2\frac{M_0}{M}\left|\frac{df}{dM}\right|dM
\:.
\end{equation}
Of course, the PS MDF includes neither the effect of the spatial
correlation of fluctuations nor that of the density excess in a
precluster region.

\begin{figure}\epsscale{0.45}
\plotone{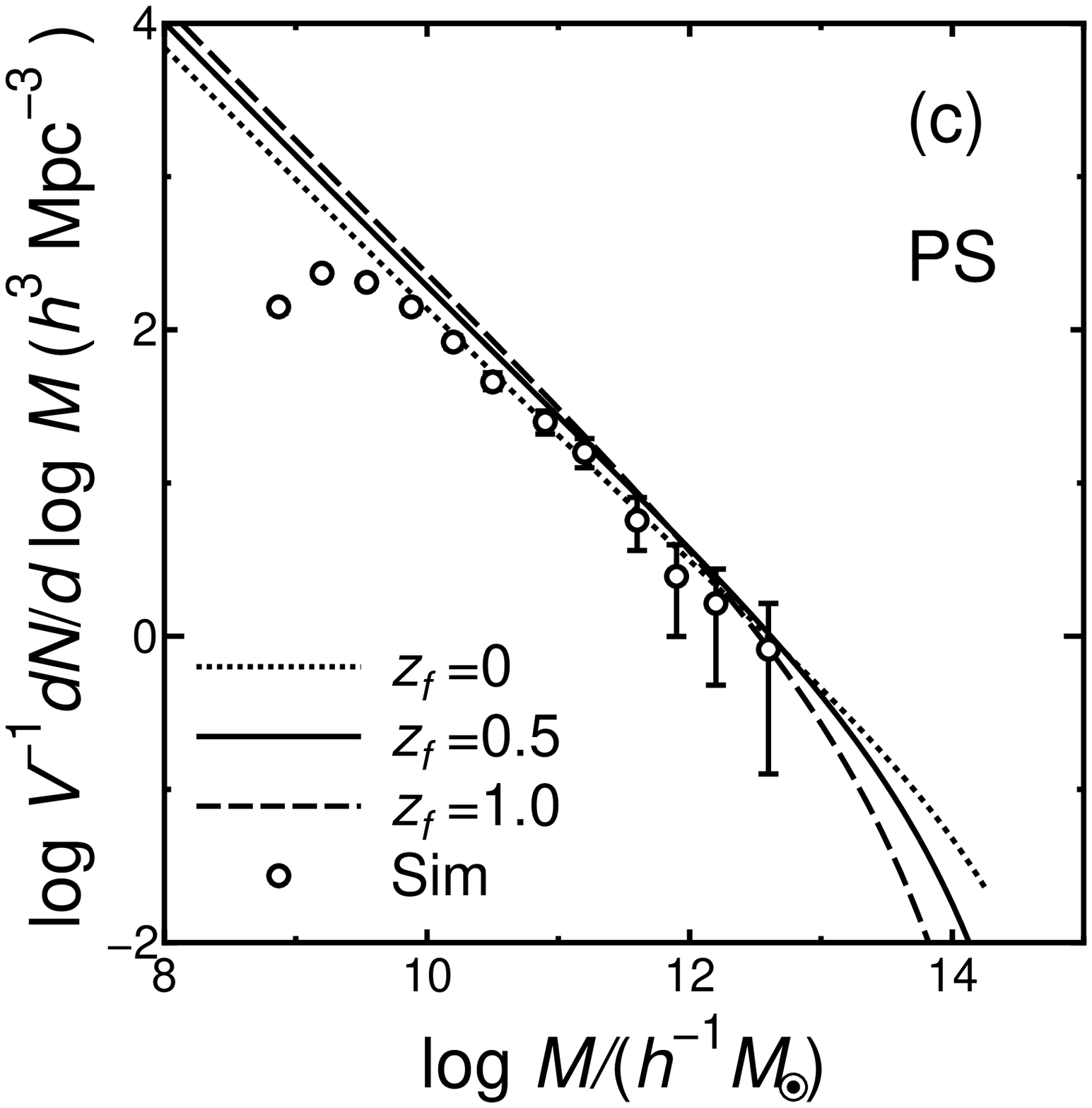} \caption{Subhalo mass distribution functions for $z_f=0$ or
0.1 (dotted lines), $z_f=0.5$ (solid lines), and $z_f=1.0$ (dashed
lines). (a) $n_{\rm SPS}$, (b) $n_{\rm EPS}$, and (c) $n_{\rm PS}$. The
MDF at $z=0$ obtained from the numerical simulation by \citet{ghi00}
is shown by the circles. \label{fig:nm}}
\end{figure}

Figure~\ref{fig:nm} shows the three MDFs at $z=z_0=0$. For the vertical
axes, we use
\begin{equation}
 \frac{1}{V}\frac{dN}{dM}=\frac{n_i}{V}\:,
\end{equation}
where $V=4\pi R_c^3/3$ and $i=$ SPS, EPS, and PS following
\citet{ghi00}. We use capital $N$ for cumulative numbers from now
on. The cluster radius, $R_c$, is the same for all the three models and
is given by $R_c=(3M_0/4\pi\rho_0)^{1/3}$, where $\rho_0$ is the average
density of the cluster and is 178 times the critical density of the
universe for our cosmological parameters.  In equation (\ref{eq:fPS}),
we use $\delta_c=\delta_c(z_f)$.

As the effective formation redshifts of a cluster, we use $z_f=0$, 0.5,
and 1 for $n_{\rm SPS}$ and $n_{\rm PS}$. 
On the other hand, if we chose $z_f=0$ in the EPS MDF, we would find
trivially that $n_{\rm EPS}=0$ except at $M=M_0$.  (That is, there are
no subhalos except for the cluster itself.)  Thus, for the EPS MDF we
show the results for $z_f=0.1$, 0.5, and 1.  The mass distribution
function $n_{\rm SPS}$ increases with $z_f$ for smaller $M$, which is
also true of $n_{\rm EPS}$ and $n_{\rm PS}$.  Figure~\ref{fig:nm} also
shows that $n_{\rm EPS}<n_{\rm SPS}<n_{\rm PS}$ for $z_f\sim 0 = z_0$
and $M\lesssim 10^{12}\; M_{\rm \sun}$.  This can be understood as
follows: When $z_f\ll 1$ (or $\delta_{c}[z_f]\approx \delta_{c}[z_0]$),
$M\ll M_0$ (or $\epsilon[0]\ll 1$), and $\delta_c(z_f)/\sigma(M)\ll 1$,
we obtain
\begin{equation}
 \beta(r) \approx  
\frac{\delta_{c}(z_f)}{\sigma(M)}
\left[1-\frac{\delta_{c}(0)}{\delta_{c}(z_f)}\alpha(r)\right]\:,
\end{equation}
and
\begin{equation}
 \beta' \approx  
\frac{\delta_{c}(z_f)}{\sigma(M)}
\left[1-\frac{\delta_{c}(0)}{\delta_{c}(z_f)}\right]\:.
\end{equation}
This implies that
\begin{equation}
 \beta'<\beta<\delta_c(z_f)/\sigma(M)\ll 1 \: ,
\end{equation}
from equations (\ref{eq:beta}) and (\ref{eq:beta'}) because
$0\leq\alpha(r)\leq 1$.
Thus, we obtain
\begin{equation}
 \left|\frac{\partial P(r,M,M_0)}{\partial M}\right|
\approx \frac{1}{\sqrt{2\pi}}
 \left[1-\frac{\delta_{c}(0)}{\delta_{c}(z_f)}\alpha(r)\right]
 \left|\frac{\partial}{\partial M}\frac{\delta_{c}(z_f)}{\sigma(M)}
\right| \:,
\end{equation}
\begin{equation}
 \left|\frac{\partial P_{\rm EPS}(M,M_0)}{\partial M}\right|
\approx \frac{1}{\sqrt{2\pi}}
 \left[1-\frac{\delta_{c}(0)}{\delta_{c}(z_f)}\right]
 \left|\frac{\partial}{\partial M}\frac{\delta_{c}(z_f)}{\sigma(M)}
\right| \:,
\end{equation}
\begin{equation}
 \left|\frac{\partial f[\geq\delta_c(z_f),M]}{\partial M}\right|
\approx \frac{1}{\sqrt{2\pi}}
 \left|\frac{\partial}{\partial M}\frac{\delta_{c}(z_f)}{\sigma(M)}
\right| \:.
\end{equation}
from equations (\ref{eq:probr1}), (\ref{eq:nocorr}), and
(\ref{eq:fPS}). Since the relation
\begin{equation}
 \left|\frac{\partial P_{\rm EPS}(M,M_0)}{\partial M}\right|
<\left|\frac{\partial P(r,M,M_0)}{\partial M}\right|
<\left|\frac{\partial f[\geq\delta_c(z_f),M]}{\partial M}\right|
\end{equation}
always holds for $0<r<R$, we have
\begin{equation}
\label{eq:N}
 n_{\rm EPS}<n_{\rm SPS}<n_{\rm PS}
\end{equation}
from equations (\ref{eq:yng}). (\ref{eq:c}), and
(\ref{eq:PS2}). Relation (\ref{eq:N}) reflects the fact that in the
model of \citet{yan96} the critical density threshold of the collapse of
the cluster is effectively between zero and $\delta_c(0)$ at $0<r<R$. On
the other hand, when $n_{\rm EPS}$ is derived, it is implicitly assumed
that the critical density threshold is uniform and is $\delta_c(0)$ in
the precluster region.  These results show that at least $n_{\rm EPS}$
and $n_{\rm PS}$ are not appropriate to describe the MDF of subhalos
especially when $z_f\lesssim 1$.

On the other hand, when $z_f\gg 1$ (or $\delta_{c}[z_f]\gg 
\delta_{c}[z_0]=1.69$), $M\ll M_0$ (or $\epsilon[0]\ll 1$), we obtain
\begin{equation}
 \beta(r)\approx \beta' \approx \delta_c(z_f)/\sigma(M)\:,
\end{equation}
from equations (\ref{eq:beta}) and (\ref{eq:beta'}).  This means that
$n_{\rm EPS}\approx n_{\rm SPS}\approx n_{\rm PS}$.  Figure~\ref{fig:nm}
shows that this relation is satisfied even when $z_f\sim 1$ for
$M\lesssim 10^{12}\; M_{\rm \sun}$.

We compare the above results with those of ultra-high resolution
numerical simulations done by \citet{ghi00}. In their simulation, a
cluster contains $\sim 5$ million particles within the final virial
radius and is simulated using a force resolution of 1.0~kpc (0.05\% of
the virial radius). The particle mass is $1.1\times 10^8\; M_{\sun}$.
Of the three MDFs $n_{\rm SPS}$, $n_{\rm EPS}$, and $n_{\rm PS}$, only
$n_{\rm SPS}$ includes both the effects of the spatial correlation of
fluctuations and the density excess in a precluster region, so we
concentrate on $n_{\rm SPS}$.  Figure~\ref{fig:nm}a shows that the
result of numerical simulation at $z=0$ is consistent with $n_{\rm SPS}$
at $z_f\sim 0.5$. The change of the MDF slope of the simulated subhalos
at $M\lesssim 10^{9}\: M_{\sun}$ seems to be caused by incompleteness
due to limited resolution.  The formation redshift, $z_f$, is reasonable
because the cluster in \citet{ghi00} does not undergo major mergers
since $z=0.5$ and the virial mass increases only by $\lesssim 30$\%
since that epoch. In fact, \citet{ghi00} compared the MDF at $z=0$ with
that of subhalos at $z=0.5$ within the {\it same physical volume}, as
determined by the virial radius of the cluster at $z=0$, and found that
they are almost identical. Thus, the MDF is ``frozen in time'' since
$z=0.5$, and $z_f$ should be $\ga 0.5$. Note that since the comparison
was done within the same physical volume, the accretion of halos from
outside cluster does not affect the evolution of the MDF. Simulation of
clusters that collapse more recently ($z_f\sim 0$) would be useful to
confirm whether $n_{\rm SPS}$ is actually superior to $n_{\rm EPS}$ and
$n_{\rm PS}$, because the three MDFs are more disparate at smaller $z_f$
(Figure~\ref{fig:nm}).  Moreover, a recently formed cluster, the effects
we have not considered (e.g., tidal stripping in the cluster) would not
have much time to operate.  One can estimate the formation redshift
$z_f$ of the simulated cluster by finding the redshift at which the MDF
(or VDF; see next section) stop evolving.

\begin{figure}\epsscale{0.45}
\plotone{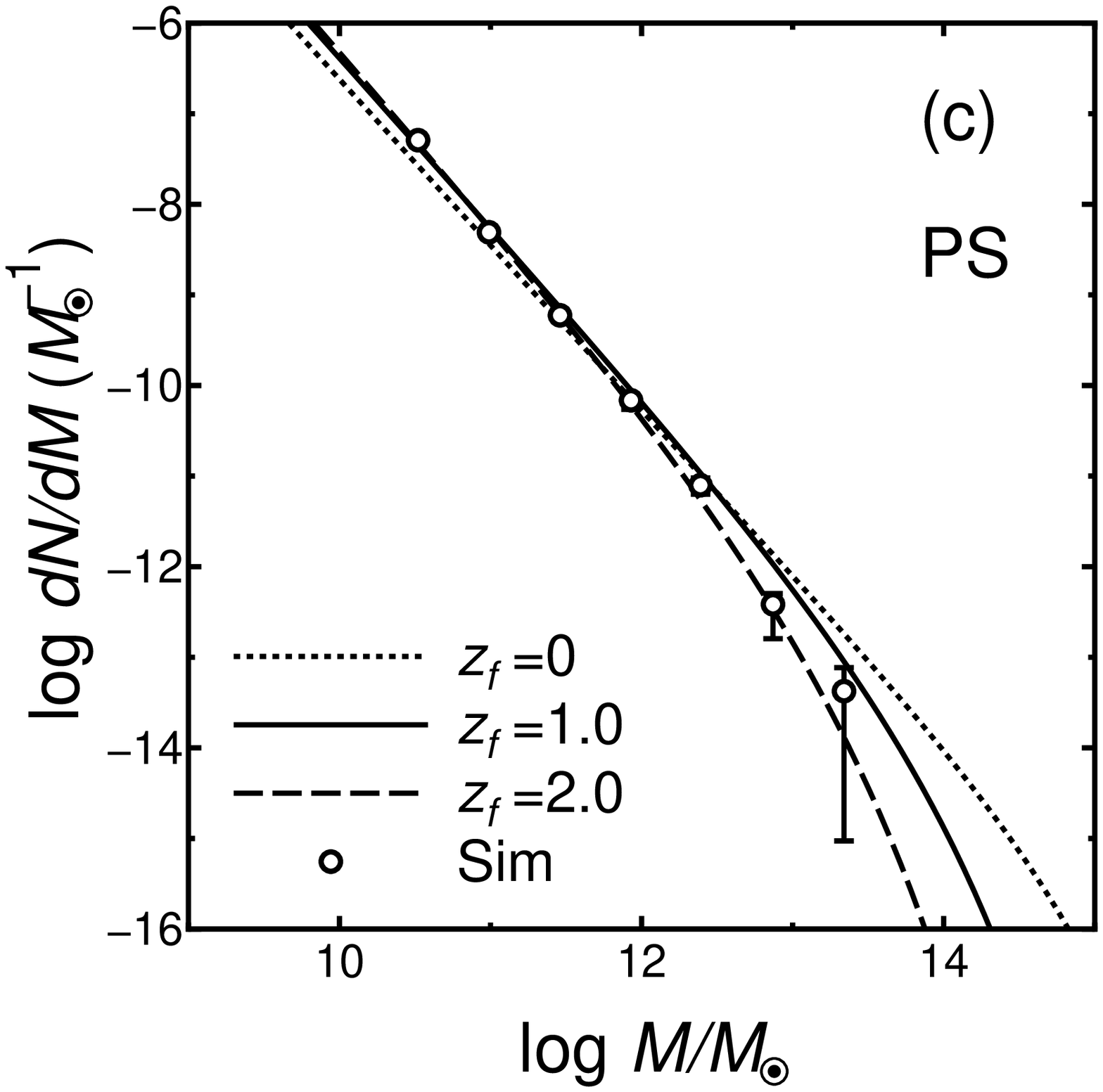} \caption{Subhalo mass distribution functions for $z_f=0$ or
0.1 (dotted lines), $z_f=1.0$ (solid lines), and $z_f=2.0$ (dashed
lines). (a) $n_{\rm SPS}$, (b) $n_{\rm EPS}$, and (c) $n_{\rm PS}$. The
MDF at $z=0$ obtained from the numerical simulation by \citet{oka99}
is shown by the circles. \label{fig:nmo}}
\end{figure}

We also compare our model with another numerical simulation done by
\citet{oka99}. They adopted cosmological parameters of $\Omega_0=1$,
$\lambda=0$, $\sigma_8=2/3$, $\Gamma=0.5$, and $h=0.5$. The present mass
of a cluster is $M_0=9.3\times 10^{14}\; M_{\sun}$. In their simulation,
the cluster contains one million particles within the final virial
radius and is simulated using a force resolution of 5.0~kpc (0.2\% of
the virial radius). The particle mass is $1.08\times 10^9\; M_{\sun}$.
Thus, the resolution is a little worse than \citet{ghi00}.
Figure~\ref{fig:nmo} shows the MDF of subhalos in the cluster. For
vertical axes, we use $dN/dM=n_i$ ($i=$SPS, EPS, and PS) following
\citet{oka99}. For $z_f\gtrsim 1$, the three models are not very
different for $M\lesssim 10^{12}\; M_{\sun}$, which is the same as in
Figure~\ref{fig:nm}. The result of \citet{oka99} at $z=0$ is also shown
in Figure~\ref{fig:nmo}. As can be seen, $n_{\rm SPS}$ at $z_f\sim 2$ is
consistent with the MDF of subhalos obtained by \citet{oka99}. In fact,
they indicated that the MDF does not evolve significantly from $z=2$ to
the present; this cluster seems to have formed earlier than that in
\citet{ghi00}.  Okamoto \& Habe argue that the reason why the MDF in
their simulated cluster agrees better with $n_{\rm PS}$ than with
$n_{\rm EPS}$ at $z_f\sim 2$ may be because of their halo-finding
algorithm. However, our results suggest this difference may be because
$n_{\rm EPS}$ does not include the effect of spatial correlations of
fluctuations and because $n_{\rm PS}$ is almost the same as $n_{\rm
SPS}$ at $z_f\sim 2$ (Figure~\ref{fig:nmo}).

\section{Tidal Stripping and Dynamical Friction}
\label{sec:tidal}

We should note that we do not consider some mechanisms that could change
the MDF of subhalos in a cluster after the cluster formation; these
mechanisms might have been effective even in cluster progenitors.

First, tidal stripping reduces the mass of subhalos. We assume that the
density profiles of subhalos and the main cluster are represented by the
so-called NFW profile:
\begin{equation}
 \rho(r)=\frac{\rho_0 r_s^3}{r(r+r_s)^2}\:,
\end{equation} 
where $r_s$ and $\rho_0$ are the characteristic radius and density of
the halo, respectively \citep*{nav97}. The mass profile is given by
\begin{equation}
 M(r) = M_{\rm vir}f(x)/f(C) \:,
\end{equation}
where
\begin{equation}
 f(x) = \ln(1+x)-\frac{x}{1+x}\;, \; \; x=\frac{r}{r_s} \:,
\end{equation}
and
\begin{equation}
\label{eq:conc}
 C \equiv \frac{r_{\rm vir}}{r_s} 
\approx 124\left(\frac{M_{\rm vir}}{1 h^{-1}\; M_{\sun}}\right)^{-0.084}
\end{equation}
\citep{kly99}. In this section, we often refer to subhalos and the main
cluster as halos, and represent the total mass of halos by their virial
masses $M_{\rm vir}$; their virial radii are represented by $r_{\rm
vir}$. The shape of the profile is characterized by the concentration
parameter $C$. The circular velocity of a halo at a radius $r$ is
defined as $v_c(r)=[GM(<r)/r]^{1/2}$. For the NFW profile, the maximum
of the circular velocity occurs at $r_{\rm max}\approx 2 r_s$. and it is
given by,
\begin{equation}
 v_{\rm peak} = \left[\frac{G M_{\rm vir}}{r_s}
\frac{f(2)}{2f(C)}\right]^{1/2} \;.
\end{equation}

The tidal radius $r_t$ of a subhalo with mass $M(r)$ and peak circular
velocity $v_{\rm peak}$ moving at a radius $R$ from the center of a
cluster with mass $M_0(R)$ and peak circular velocity $v_0$ is given by
the radius at which the gravity force of the subhalo is equal to the
tidal force of the main cluster. The radius can be derived by solving
the following equation for the tidal radius $r_t$:
\begin{equation}
 \left(\frac{R}{r_t}\right)^3 \frac{M(r_t)}{M_0(R)}
= 2-\frac{R}{M_0}\frac{\partial M_0}{\partial R} \;,
\end{equation}
which is equivalent to 
\begin{equation}
 \frac{f(x_t)}{f(x_R)}=\left(\frac{x_t}{x_R}\right)^3
\left(\frac{r_s v_0}{R_s v_{\rm peak}}\right)^2
\left[2-\frac{x_R^2}{(1+x_R)^2 f(x_R)}\right]\;,
\end{equation}
where $x_t=r_t/r_s$, $x_R=R/R_s$, and $R_s$ is scale radius of the
cluster \citep{kly99}.

On the other hand, tidal stripping can occur by resonances between the
force the subhalo exerts on the dark matter particle and the tidal force
by the main cluster. In this case, the tidal radius can be obtained by
solving the following equation for the tidal radius:
\begin{equation}
 \frac{G M(r_t)}{r_t^3} = \frac{G M_0(R)}{R^3} \;,
\end{equation}
which is equivalent to 
\begin{equation}
 \frac{f(x_t)}{f(x_R)}=\left(\frac{x_t}{x_R}\right)^3
\left(\frac{r_s v_0}{R_s v_{\rm peak}}\right)^2
\end{equation}
\citep{kly99}. We take the smaller of the two estimates of $r_t$.

\begin{figure}\epsscale{0.45}
\plotone{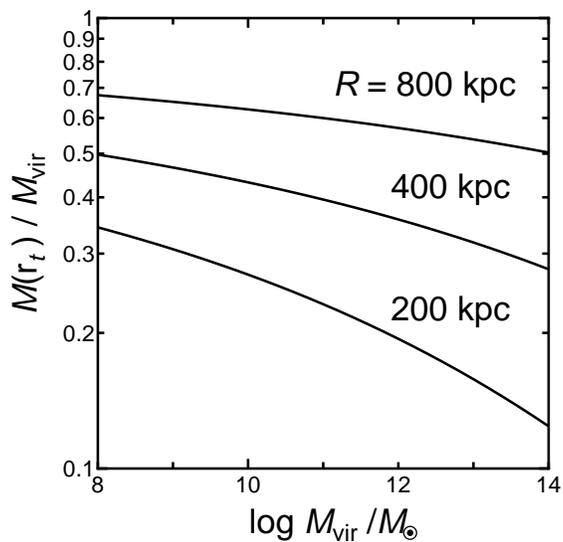} \caption{The fraction of mass within the tidal radius
of subhalos with different masses for several distances from the cluster
center ($R$). The cosmological and cluster parameters are the same as
those in Figure~\ref{fig:nm}. \label{fig:tidal}}
\end{figure}

Figure~\ref{fig:tidal} shows the mass fraction of a subhalo within the
tidal radius as a function of the `subhalo' virial mass $M_{\rm
vir}$. The cosmological and cluster parameters are the same as those in
Figure~\ref{fig:nm}. As can be seen, tidal stripping is very effective
when the distance from the cluster center is small. Moreover, it is
effective when $M_{\rm vir}$ is large, because the concentration
parameter $C$ is small (equation~[\ref{eq:conc}]). In a real cluster,
the effect of tidal stripping depends on the orbit of a subhalo in a
cluster. Since it is difficult to discuss the orbits of subhalos
analytically, numerical simulations are required to estimate the effect
of tidal stripping on entire subhalos in a cluster quantitatively.
\citet{ghi98} estimated that for a circular velocity larger than $\sim
10^{2.2}\;\rm km\; s^{-1}$ ($M_{\rm vir}\sim 2.5\times 10^{11}\;
M_{\sun}$), the masses of subhalos are smaller than those of isolated
halos by $\sim 20-50$\% on the average due to tidal stripping. Thus, the
tidal stripping would affect the comparison between our numerical model
and the results of numerical simulations done in \S\ref{sec:MDF},
especially for the high mass end of the MDFs.

Second, the most massive subhalos often merge with the central ``cD
galaxy'' via dynamical friction. Thus, we may overestimate the number of
these subhalos. The time-scale of dynamical friction can be estimated by
using Chandrasekhar's formula. When the density distribution of a
cluster is approximated by an isothermal density distribution, a subhalo
with the initial position of $R$ reaches the cluster center in the time
of
\begin{eqnarray}
 t_{\rm fric}&=&\frac{1.17}{\ln \Lambda}
        \frac{R^2 v_0}{G M(r_t)}\\
    &=& 6.4\times 10^9 {\rm\; yr}\left(\frac{\ln \Lambda}{5}\right)^{-1}
\left(\frac{R}{1\:\rm Mpc}\right)^2
\left(\frac{v_0}{1200\rm\: km\: s^{-1}}\right)
\left[\frac{M(r_t)}{10^{13}\: M_{\sun}}\right]^{-1}\;, \label{eq:fric}
\end{eqnarray}
where $\Lambda=R_c M_0(R)/[R M(t_t)]$, and $R_c$ is the cluster radius
\citep{bin87}. Note that the time-scale (equation [\ref{eq:fric}]) is
not much different from the one obtained by \citet{kly99} without the
approximation of an isothermal distribution.

\begin{figure}\epsscale{0.45}
\plotone{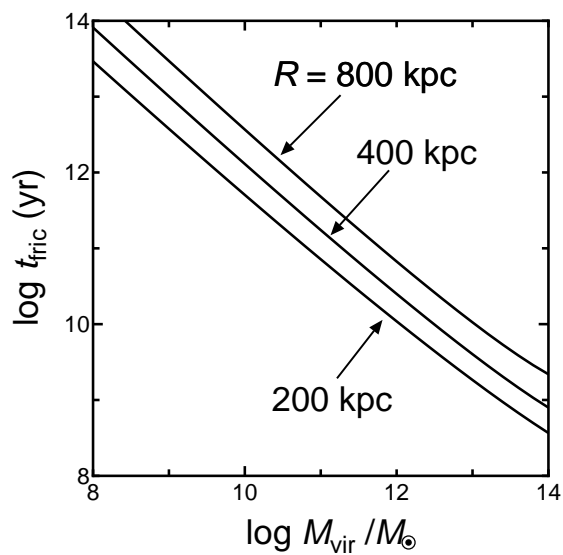} \caption{The dynamical friction time for different
 masses of subhalos for several distances from the cluster center
 ($R$). The cosmological and cluster parameters are the same as those in
 Figure~\ref{fig:nm}.\label{fig:fric}}
\end{figure}

Figure~\ref{fig:fric} shows the time-scale of dynamical friction $t_{\rm
fric}$ as a function of the subhalo virial mass. The cosmological and
cluster parameters are the same as those in Figure~\ref{fig:nm}. Massive
subhalos near the cluster center are most affected by dynamical
friction. The fraction of subhalos that fall into the cluster center
depends on the distribution of their orbits. In the numerical
simulations done by \citet{ghi00}, 60\% of subhalos with the circular
velocity of $\gtrsim 300\;\rm km\: s^{-1}$ ($M_{\rm vir}\gtrsim 2.5
\times 10^{12}\; M_{\sun}$) have merged into the central cD galaxy since
$z=3$, although the mergers have been less effective after $z=1$. Thus,
the MDFs derived in \S\ref{sec:MDF} should be compared with the results
of numerical simulations and observations with caution for those massive
subhalos.

On the other hand, the distribution of peak circular velocities $v_{\rm
peak}$ is less affected by tidal stripping, because the removal of the
outer parts of a subhalo does not affect $v_{\rm peak}$ very
strongly. \citet{kly99} showed that even if $\sim 70$\% of the initial
mass of a subhalo is removed by tidal stripping, $v_{\rm peak}$ changes
only by $\sim 20$\%. Since the numerical simulations done by
\citet{ghi98} showed that tidal stripping reduces the mass of massive
subhalos at most by $\sim 50$\%, the tidal stripping may not affect the
distribution function of values of $v_{\rm peak}$ (VDF) for subhalos
significantly. Thus, in the next section, we investigate the VDF for
subhalos, although the high-velocity end of the VDF may still be
affected by dynamical friction, or mergers of subhalos. Observationally,
$v_{\rm peak}$ can be obtained more easily than $M$ through the velocity
dispersion of stars or the rotation curve of a galaxy.  Thus, the VDF is
more useful for comparison with observations than the MDF.

\section{Velocity Distribution Function}
\label{sec:VDF}

\begin{figure}\epsscale{0.45}
\plotone{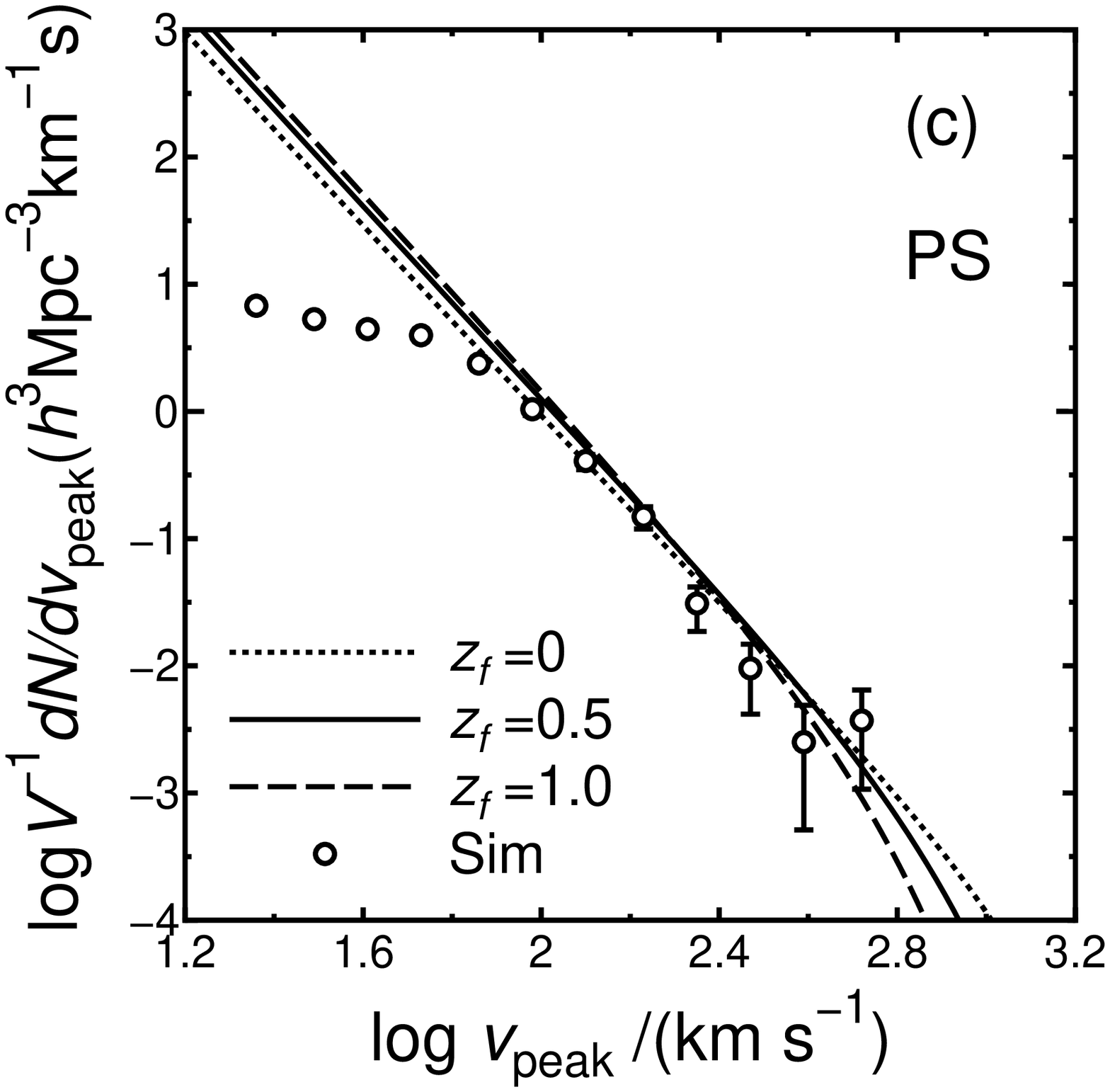} \caption{Velocity distribution functions for $z_f=0$
or 0.1 (dotted lines), $z_f=0.5$ (solid lines), and $z_f=1.0$ (dashed
lines). (a) $n_{\rm SPS}$, (b) $n_{\rm EPS}$, and (c) $n_{\rm PS}$. The
VDF at $z=0$ obtained from the numerical simulation by \citet{ghi00}
is shown by the circles. \label{fig:nv}}
\end{figure}

In order to derive the VDF of subhalos from the MDF, we need the
relation between the velocity and mass. Since there is no simple
analytically derived relation, we use the relation derived by numerical
calculations for isolated halos. \citet*{nav97} found that the relation
between $v_{\rm peak}$ and $M$ is
\begin{equation}
\label{eq:VM}
 v_{\rm peak}=1510\;{\rm km\; s^{-1}}
\left(\frac{M}{10^{15}\; M_{\sun}}\right)^{1/3.28}
\end{equation}
for the Einstein de-Sitter universe. Although the relation is for $z=0$,
it is not much different for $z\lesssim 2$ \citep{ghi98,bul01}. We
assume that this relation holds for subhalos, at least before the
collapse of the cluster at $z=z_f$. Using equation (\ref{eq:VM}), we
transform the MDFs derived in \S\ref{sec:MDF} into VDFs; the results
are shown in Figure~\ref{fig:nv}. Note that when we discuss the VDFs,
$n_i$ ($i=$ SPS, EPS, and PS) is a function of $v_{\rm peak}$, and
\begin{equation}
 \frac{1}{V}\frac{dN}{dv_{\rm peak}}=\frac{n_i}{V}\;.
\end{equation}
Since $v_{\rm peak}$ is a monotonically increasing function of $M$, the
relation among $n_{\rm SPS}$, $n_{\rm EPS}$, and $n_{\rm PS}$ is the
same as that for MDF.  In Figure~\ref{fig:nv}, the results of numerical
simulations are also shown \citep{ghi00}. In general, $n_{\rm SPS}$ at
$z_f=0.5$ is consistent the result of numerical calculation as was shown
to be the case for the MDF.  This may mean that tidal stripping does not
significantly affect the MDF of subhalos in the simulated cluster. The
deviation of the simulated cluster data from the model at $v_{\rm
peak}\lesssim 10^{1.9}\rm\; km\; s^{-1}$ is caused by finite numerical
resolution. The slope of the VDF in the simulated cluster is a little
steeper than that of $n_{\rm SPS}$ for $v_{\rm peak}\gtrsim
10^{2.3}\rm\; km\; s^{-1}$ if we remove the largest $v_{\rm peak}$
point.  This may be caused by mergers of massive subhalos with the
central cD galaxy.

If the VDF is not significantly influenced by tidal stripping, we can
use it to derive the effective formation redshift of a
cluster. Figure~\ref{fig:virgo} shows the observed cumulative number of
subhalos in the Virgo cluster derived from galaxy data
\citep{bin85,moo99} and the integrated $n_{\rm SPS}$ for various
$z_f$. The cosmological parameters for $n_{\rm SPS}$ are the same as
those in \citet{ghi00}. The comparison suggests that $z_f\sim 2$,
although the slope of the Virgo data does not match that of $n_{\rm
SPS}$ for $v_{\rm peak}/v_0\lesssim 0.1$.  Here, $v_0$ is the peak
velocity $v_{\rm peak}$ of the rotation curve for applicable to the
total mass of the cluster.  The deviation for low $v_{\rm peak}$
subhalos may be because the corresponding galaxies are too faint to be
detected. Note that that the derived formation redshift, $z_f\sim 2$,
is the redshift at which the growth of subhalos stopped.  This
corresponds to the time when the potential wells of the cluster
progenitors, in which the subhalos were located, became sufficiently
deep, even if the progenitors had not yet merged into one cluster.
Since the subhalos stopped growing, the formation of galaxies in them is
also expected to decrease for $z<z_f$.  In the irregular Virgo cluster,
$z_f$ may correspond to the time at which the oldest galaxies were
formed, even though the cluster itself is still merging at present.

We note that we cannot rule out the possibility that $z_f$ is somewhat
smaller than 2 because of the ambiguity of the number and circular
velocity of massive subhalos. 
In general, for massive subhalos, dynamical friction may cause them to
merge into one large (cD) galaxy, which reduces their number. However,
although there is a moderately large galaxy (M87) at the center of the
Virgo cluster, it is not usually classified as a cD galaxy and is
probably not the most luminous galaxy in the cluster.  Alternatively,
the circular velocity of the galaxy residing in a massive subhalo may
not represent that of the massive subhalo; the former may be smaller
than the latter. In fact, \citet{dav96} showed that the temperature of
diffuse X-ray gas is systematically higher than that of kinetic
temperature of stars for bright elliptical galaxies.  If the temperature
of the X-ray gas is considered to represent the depth of potential well
of the halo around the elliptical galaxy, it shows the inconsistency of
circular velocity between galaxies and subhalos.  Moreover,
\citet{mat01} showed that bright elliptical galaxies with extend X-ray
emission tend to have high X-ray temperature compared to those with
compact X-ray emission for a given stellar velocity dispersion.  On the
other hand, the X-ray gas may be affected by other heating processes
(e.g., supernovae).

These results suggest the existence of very large, massive subhalos
whose X-ray temperature is much higher than kinetic temperature of the
stars in the galaxies located in them.  Recent {\it Chandra}
observations have revealed the presence of such massive subhalos in
clusters; most of them are moving with the Mach number of $\sim 1$ in
clusters without losing their identity. Thus, in the next section, we
discuss the relation between the large X-ray subhalos and VDF.

\section{Cold Fronts}
\label{sec:coldf}

Large X-Ray subhalos are found in some clusters as low temperature X-ray
components with cold fronts \citep{mar00,vik01,maz01,maz02,sun02}.  The
cold fronts are contact interfaces between the cooler gas in the subhalo
and the hotter intracluster medium in the main cluster.  The ratio of
X-ray temperature of the subhalos to that of their host clusters is
about $1/2$.  Thus, if the X-ray temperature is proportional to $v_{\rm
peak}^2$, the corresponding velocity ratio is $v_{\rm peak}/v_0\sim
1/\sqrt{2}$.

\citet{maz01} suggested that a cluster with a large subhalo may be the
result of the collapse of two different perturbations in the primordial
density field on two different linear scales at nearly the same location
in space. As the density field evolves, both perturbations start to
collapse. The small-scale perturbation collapses first and forms a large
subhalo, and then the larger perturbation collapses and forms a
cluster. If the initial position of the subhalo was only slightly offset
from the center of the cluster, it would infall into the cluster center
with small velocity and oscillate around the cluster center. This would
prevent the cool gas in the subhalo from being removed by ram-pressure
stripping or the Kelvin-Helmholtz instability. We note that
\citet{fuj02} showed that the observed subhalos with cold fronts are
stable for $0.5-1$~Gyr against large-scale Kelvin-Helmholtz instability.

\begin{figure}\epsscale{0.45}
\plotone{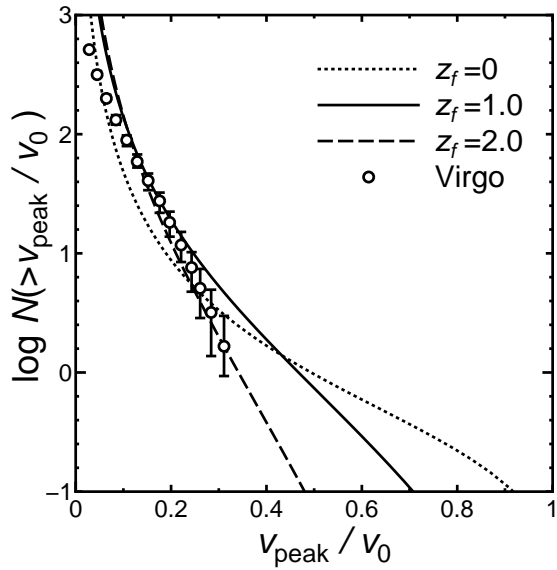} \caption{Circles show the observed cumulative number of
subhalos in the Virgo cluster derived from galaxy observations
\citep{bin85,moo99}.
The lines give the integrated SPS MDF for $z_f=0$
or 0.1 (dotted lines), $z_f=1.0$ (solid lines), and $z_f=2.0$ (dashed lines).
The integrated SPS MDF was calculated using the
cosmological parameters in \citet{ghi00}. \label{fig:virgo}}
\end{figure}

Therefore, it would be interesting to estimate the probability of
finding such large subhalos near centers of clusters, and to compare
this probability with the observed rate of occurrence of cold fronts.
If we had enough observational data about the massive subhalos, these
could be used to study the validity of our MDF or VDF model.  However,
at present the number of clusters in which large X-ray subhalos are
found is too small to allow a detailed discussion of the rate of
occurrence of cold fronts.  Thus, in this study, we just predict the
nature of the massive subhalos, assuming that our model is correct. The
cumulative number of the most massive halos is larger for smaller $z_f$
(Figure~\ref{fig:virgo}), because hierarchical clustering proceeds
furthest. Thus, we use $n_{\rm SPS}$ at $z_f=0$ to estimate the maximum
probability of finding those subhalos.

\begin{figure}\epsscale{0.45}
\plotone{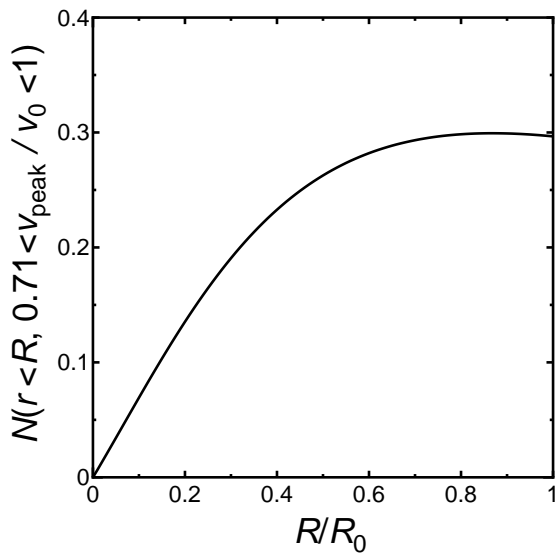} \caption{The average number of subhalos with
$0.71<v_{\rm peak}/v_0<1$ within $r<R$. We assume that
$z_f=0$. \label{fig:rad}}
\end{figure}

Figure~\ref{fig:virgo} shows that the cumulative number, $N(0.71<v_{\rm
peak}/v_0<1)$, is about 0.3, which means that at most one third of
clusters should have large subhalos, some of which may have cold fronts
similar to the observed ones. In order to find the initial position of
the subhalos in a precluster region, we calculate the number by changing
the radius, $R$, within which the conditional probability (equation
[\ref{eq:prob}]) is averaged. Figure~\ref{fig:rad} shows the result for
the same cluster in Figure~\ref{fig:virgo}. Since $N(r<R, 0.71<v_{\rm
peak}/v_0<1)$ is almost constant for $R/R_0\gtrsim 0.5$, it is most
likely that the initial position of the subhalo is $R/R_0\lesssim 0.5$.
Because we do not consider the spatial distribution of mass in a
precluster region and the detailed evolution of cluster collapse, we
cannot exactly predict the infall or oscillating velocity of the subhalo
in a cluster. However, if the initial position of the subhalo is well
inside the precluster region, the infall or oscillating velocity may be
subsonic or at most transonic, which may be consistent with the
observations \citep{mar00,vik01,maz01,maz02}. Note that if the initial
position of a subhalo is near $r=R_0$, the Mach number of the infall
velocity should be $\sim 2-3$ \citep{sar02}. The above result suggests
that most of the X-ray substructures in clusters have not fallen in from
large distances, as would be true for cluster mergers.  Instead, most of
the cold fronts may be due to internal subhalos.  The fact that clusters
having subhalos moving with highly supersonic velocities
\citep[e.g. 1E0657$-$56;][]{mar02} are rarely found may support this
idea, although the number of clusters for which the velocities of
subhalos are derived is small.

As is mentioned in \S\ref{sec:tidal}, the large subhalos should be
affected by dynamical friction ($v_c\gtrsim 300\;\rm km\: s^{-1}$
according to \citealt{ghi98}). In approximately the time shown in
equation~(\ref{eq:fric}) the subhalos reaches the cluster center.
If the subhalos are not completely disrupted before they reach the
cluster center, the clusters should have a potential well with two
different spatial scales. In fact, \citet{ike96} found a double dark
halo distribution in the Fornax cluster. On the other hand, in the
central regions of many clusters, cool gas is observed
\citep{fab94}. These subhalos may provide cool gas into the cluster
center.

\section{Conclusions}
\label{sec:conc}

We have constructed a model of the mass distribution function (MDF) of
subhalos in clusters, including the fact that the subhalos grew from
spatially correlated density fluctuations in high-density regions.  If a
cluster forms at a low redshift, the derived mass distribution function
of subhalos is intermediate between the Press-Schechter mass
distribution function and a conditional mass distribution function
derived by an extension of the PS formalism. However, if a cluster forms
at high redshift, our subhalo MDF is nearly identical to these other
MDFs for low mass subhalos.  The velocity distribution functions show
similar trends.  We compare our model MDF with the results of $N$-body
numerical simulations.  We find that they are consistent with each
other.  The comparison of our model with the subhalo population derived
from galaxy data suggests that the subhalos in Virgo cluster stopped
growing at $z \sim 2$.  Our model predicts that at most 1/3 of clusters
should have large X-ray subhalos like the ones observed as cold fronts
with {\it Chandra}.  Many of the observed X-ray substructures in
clusters may not be the result of cluster mergers; instead, they may be
due to internal substructures within the clusters.

\acknowledgments

We thank W. C. Saslaw for a useful discussion. We are also grateful to
K. Yoshikawa and T. Okamoto for useful comments. We acknowledge the
referee for helpful suggestions. C. L. S. was supported in part by
$Chandra$ Award Number GO1-2123X, issued by the $Chandra$ X-ray
Observatory Center, which is operated by the Smithsonian Astrophysical
Observatory for and on behalf of NASA under contract NAS8-39073, and by
NASA XMM Grant NAG5-10075.

\newpage

\begin{figure}\epsscale{1.0}
\plottwo{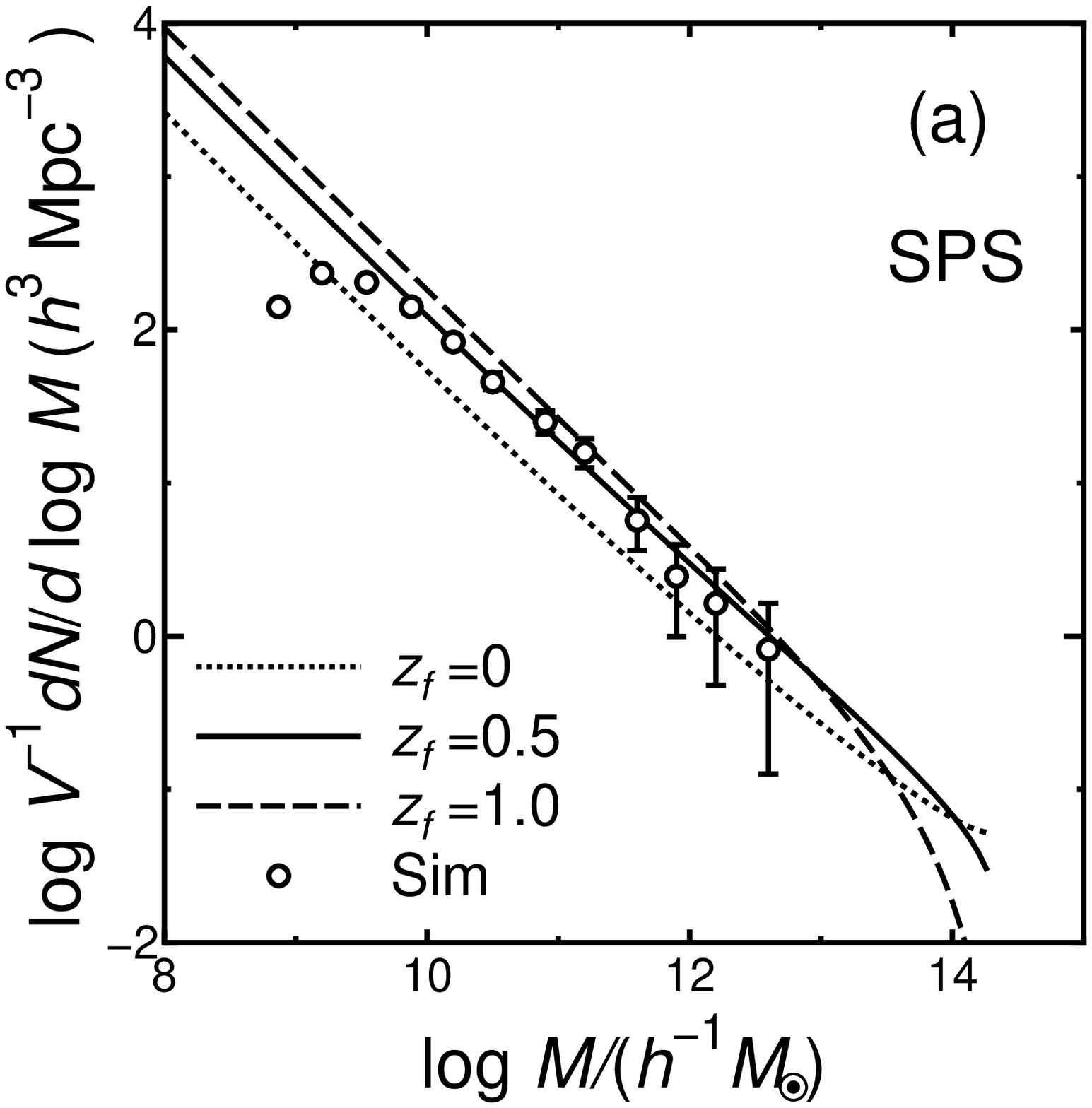}{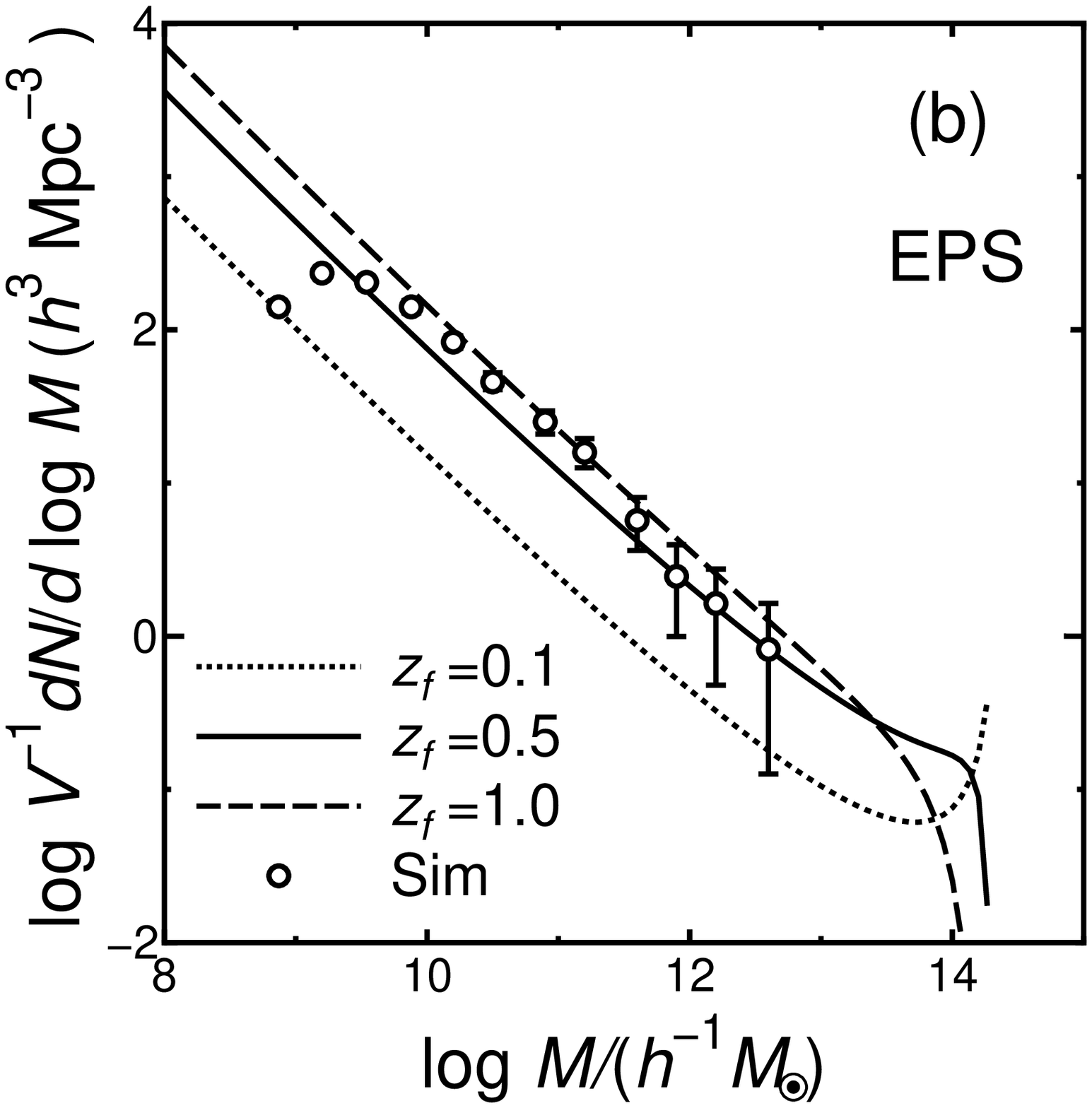}
\end{figure}

\begin{figure}\epsscale{1.0}
\plottwo{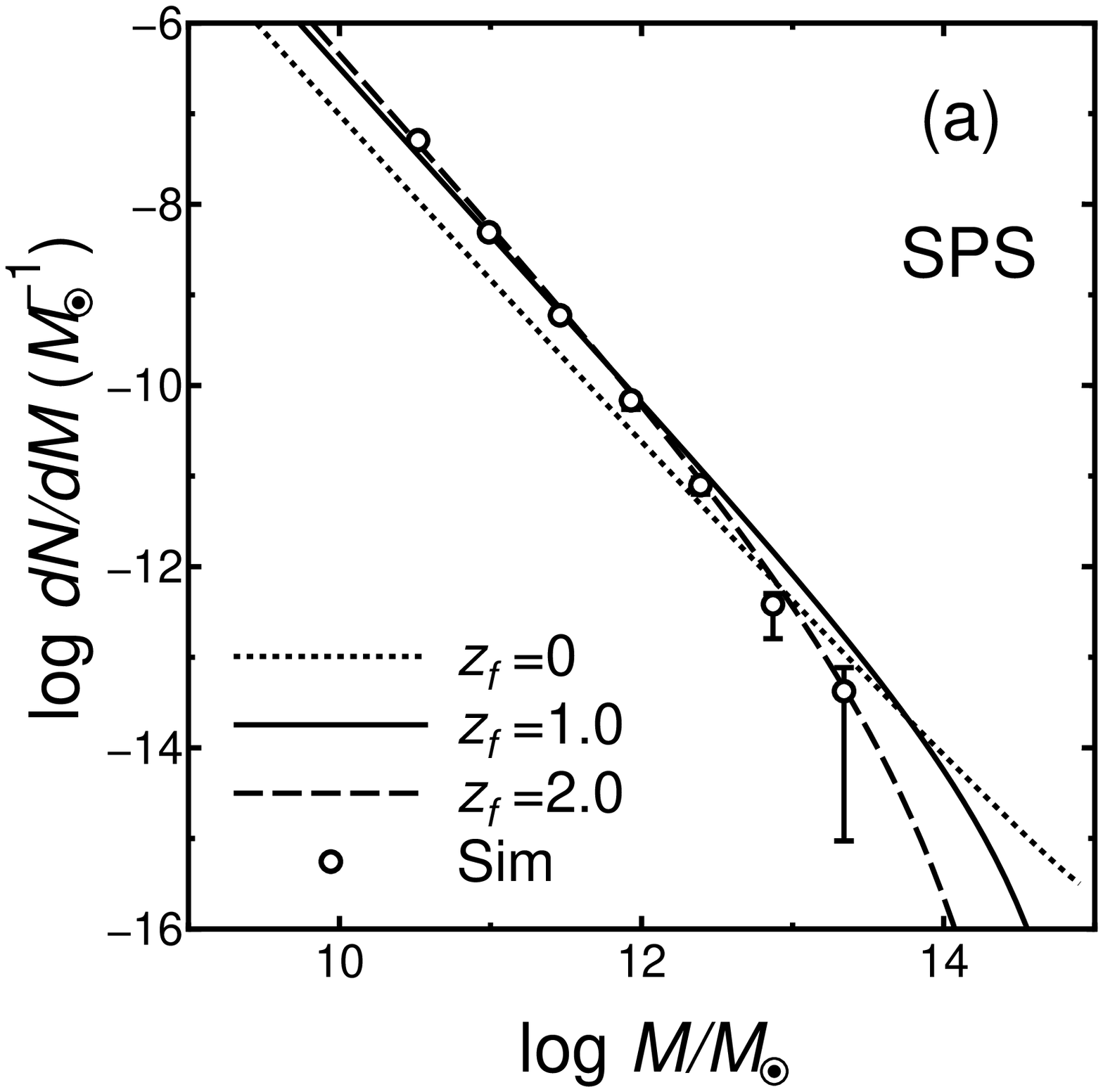}{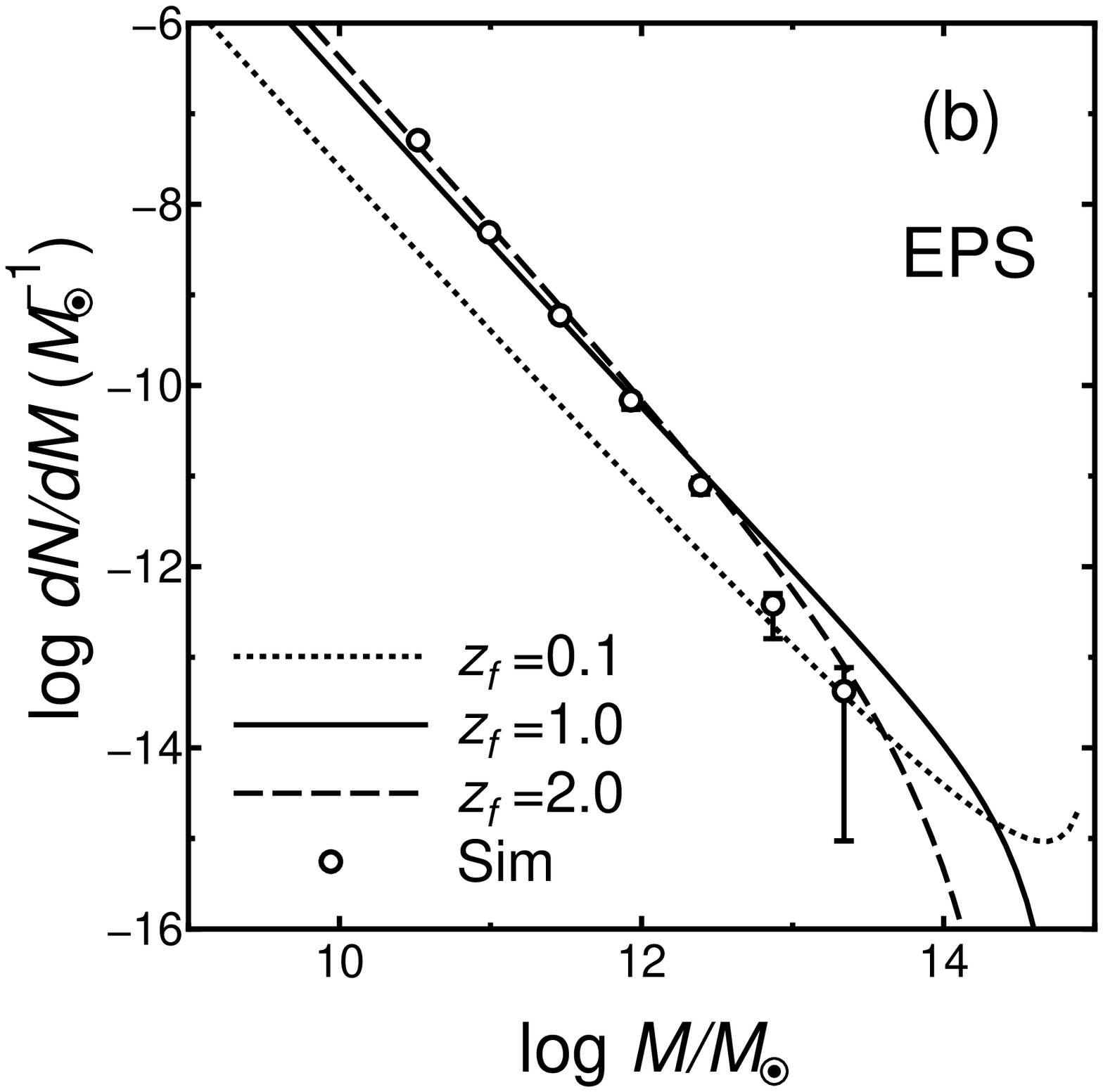}
\end{figure}

\begin{figure}\epsscale{1.0}
\plottwo{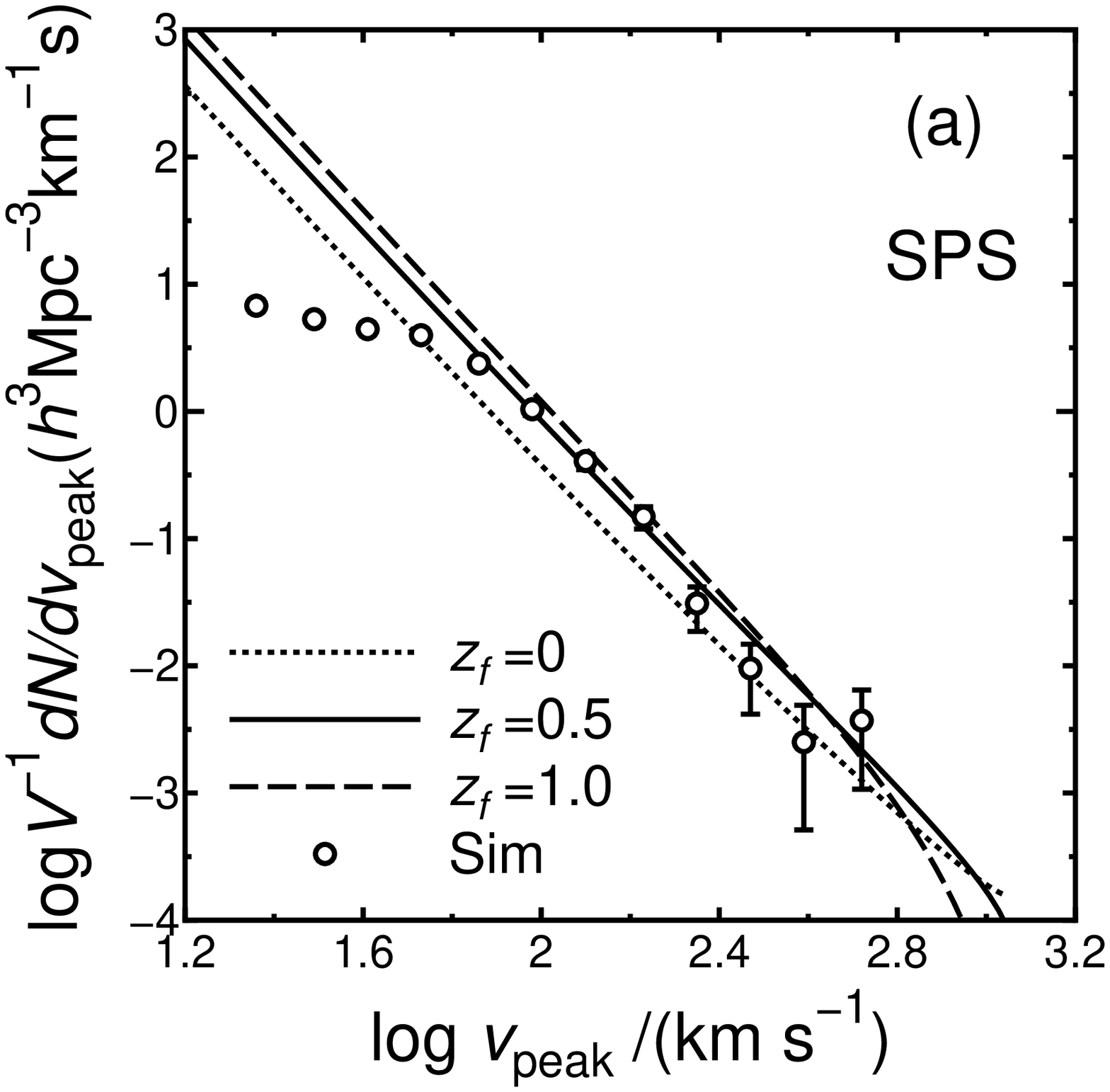}{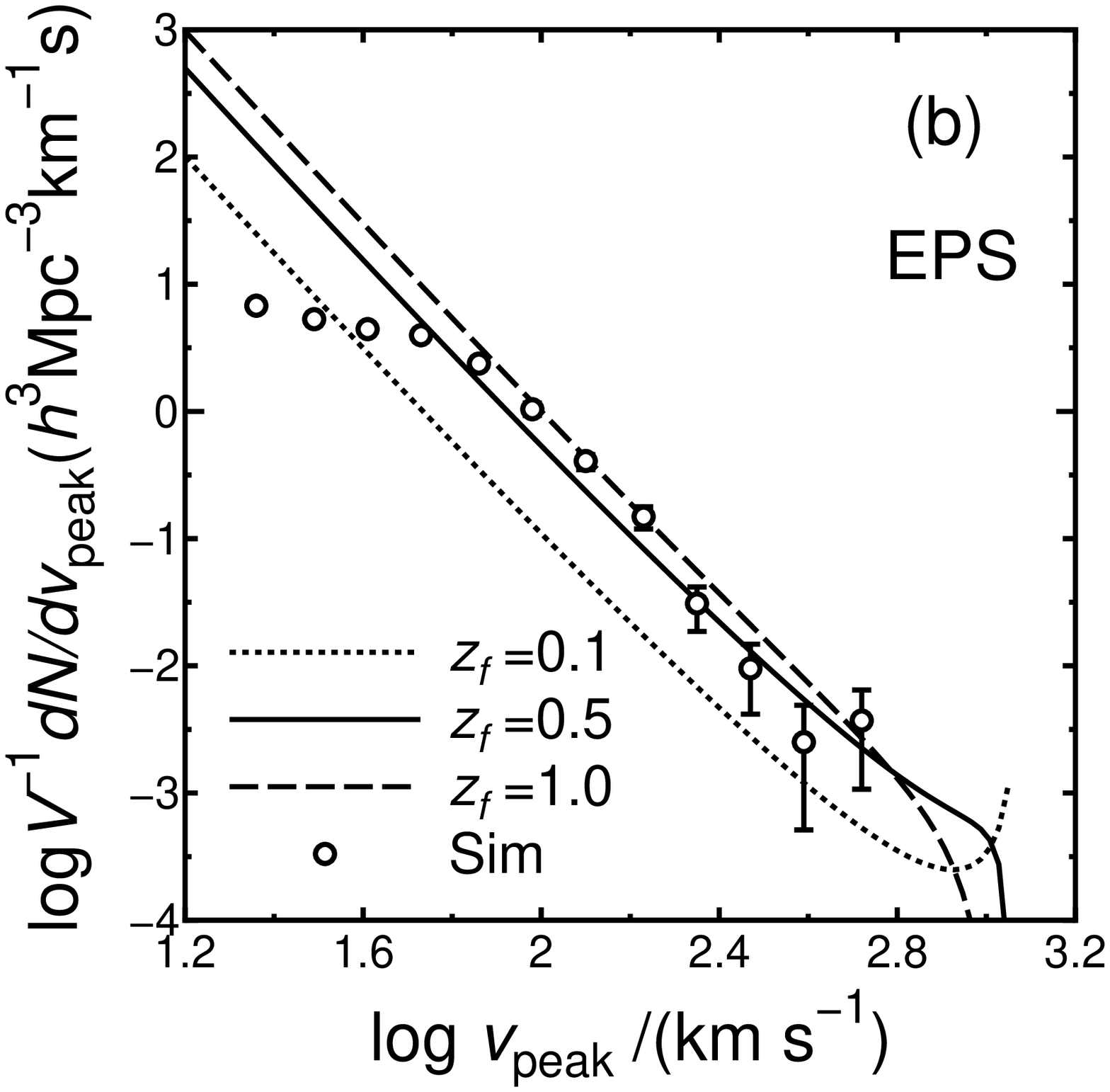}
\end{figure}

\end{document}